\documentclass[12pt]{article}
\usepackage{amsbsy,amsmath,amsthm,amsfonts,amssymb,times,subfigure,graphics,graphicx,bm,mathtools,commath}
\usepackage{multirow}
\usepackage{thmtools}
\RequirePackage{natbib}
\usepackage{algorithm,algorithmic}
\usepackage{mathrsfs}
\usepackage{caption}
\usepackage{wrapfig}
\declaretheorem[style=definition]{example}

\theoremstyle{definition}

\newcommand{\I}{{\mathrm{I}}}

\renewcommand\thmcontinues[1]{Continued}

\newcommand{\pcite}[1]{\citeauthor{#1}'s \citeyearpar{#1}}
 
\usepackage [english]{babel}
\usepackage [autostyle, english = american]{csquotes}
\MakeOuterQuote{"}

\graphicspath{{./plots/}}
\begin{document}

\title{MCMC for GLMMs}
\author{Vivekananda Roy\\
  Department of Statistics, Iowa State University, USA}
% \email{\{vroystat,lijinz\}@iastate.edu}}
%
 \date{}

\maketitle              % typeset the header of the contribution
\begin{abstract}
  Generalized linear mixed models (GLMMs) are often used for analyzing
  correlated non-Gaussian data. The likelihood function in a GLMM is
  available only as a high dimensional integral, and thus closed-form
  inference and prediction are not possible for GLMMs. Since the
  likelihood is not available in a closed-form, the associated
  posterior densities in Bayesian GLMMs are also
  intractable. Generally, Markov chain Monte Carlo (MCMC) algorithms
  are used for conditional simulation in GLMMs and exploring these
  posterior densities. In this article, we present different state of
  the art MCMC algorithms for fitting GLMMs.  These MCMC algorithms
  include efficient data augmentation strategies, as well as
  diffusions based and Hamiltonian dynamics based methods. The
  Langevin and Hamiltonian Monte Carlo methods presented here are
  applicable to any GLMMs, and are illustrated using three most
  popular GLMMs, namely, the logistic and probit GLMMs for binomial
  data and the Poisson-log GLMM for count data. We also present
  efficient data augmentation algorithms for probit and logistic
  GLMMs. Some of these algorithms are compared using a numerical
  example.
% We discuss available theoretical
% results and undertake numerical examples to compare these algorithms.
\end{abstract}

\noindent {\it Key words:} Bayesian GLMM; Data augmentation; EM; GLM; Hamiltonian Monte
Carlo; MALA; Metropolis-Hastings; Mixed models; Monte
Carlo maximum likelihood; spatial GLMM

\section{Introduction}
\label{sec:int}

Generalized linear mixed models (GLMMs) are a natural extension of
both the linear mixed models and the generalized linear models
(GLMs). GLMMs allow for non-Gaussian responses and the random effects
in the GLMMs can accommodate overdispersion often present in non-Gaussian
data, as well as dependence among correlated observations arising from
longitudinal or repeated measures studies. GLMM is one of the most
frequently used statistical models \citep{jian:thua:2007}.

% Since the likelihood function in
% GLMMs is not available in closed form, but only as a high-dimensional
% integral, the posterior density for Bayesian GLMMs is intractable
% for any choice of prior on the unknown parameters. Generally, Markov chain
% Monte Carlo (MCMC) algorithms are used for exploring these posterior densities.
%\cite{boot:hobe:1998}

Unlike the linear mixed models and the GLMs, the likelihood function
for the GLMM is not available in a closed-form. Since the likelihood
function for the GLMM is obtained by integrating a GLM likelihood with
respect to the distribution of the random effects, it is available only
as a high-dimensional integral. Both analytical and Monte Carlo
approximations for the GLMM likelihood have been proposed in the
literature. For example, \cite{bres:clay:1993} considered the
Laplace's method, and \cite{wolf:ocon:1993} used a Taylor expansion
for the integral approximation. Implementations involving Markov chain
Monte Carlo (MCMC) methods include \cite{zege:kari:1991} who used Gibbs
sampling and \cite{mccu:1997} who used Metropolis-Hastings (MH)
algorithm for making inference in GLMMs. Use of sampling-based methods
in GLMMs can be found in \cite{mccu:1994, game:1997, boot:hobe:1999} among others.

Over the last two decades, extensive efforts have gone into producing
efficient, fast mixing MCMC algorithms, in general, and in the context
of GLMMs, in particular. For example, effective data augmentation (DA)
strategies have been proposed for specific GLMMs \citep[see
e.g.][]{wang:roy:2018, pols:scot:wind:2013, wang:roy:2018a,
  rao:roy:2021}. Also novel MH algorithms based on Langevin diffusions
and Hamiltonian dynamics such as the Metropolis adjusted Langevin
algorithms (MALA) \citep{robe:twee:1996b} and the Hamiltonian Monte
Carlo (HMC) algorithms \citep{neal:2011} have now emerged as the
popular methods for MCMC sampling due to their ability to make distant
moves with high acceptance probability and favorable scalability with
respect to increasing state space dimensions. MALA and HMC have also
been applied for inference for GLMMs \cite[see e.g.][]{roy:zhan:2021,
  burk:2017}. The goal of this article is to present these efficient MCMC
algorithms for fitting GLMMs.

% whereas \cite{mccu:1997} uses importance sampling or Monte
% Carlo EM algorithm for making inference for GLMMs. Indeed For maximum
% likelihood estimation \cite{mccu:1994} \cite{mccu:1997}
% \cite{boot:hobe:1999} use iid samples based on rejection sampling and
% importance sampling to perform Monte Carlo EM.

% GLMMs are popular for analyzing different types of correlated
% observations. Using unobserved Gaussian random effects, GLMMs permit
% additional sources of variability in the data.

The rest of the article is organized as follows. In
Section~\ref{sec:lik}, we present the likelihood function for GLMMs
and describe two sampling-based approaches, namely the Monte Carlo EM
and the Monte Carlo maximum likelihood methods for approximating this
likelihood function. In Section~\ref{sec:condsim}, we construct
different MCMC algorithms for conditional simulation in GLMMs. Thus,
these algorithms can be used for sampling-based inference in GLMMs.
Several popular GLMMs are used to illustrate these algorithms.  Next,
Section~\ref{sec:bayes} presents MCMC algorithms for Bayesian GLMMs
under popular priors on the fixed effects and the variance components
parameters. Section~\ref{sec:num} contains comparisons of some of
these MCMC algorithms using a publicly available dataset. Some concluding remarks are provided in Section
\ref{sec:disc}.

\section{Likelihood function for GLMMs}
\label{sec:lik}
Let $Y = (Y_1, \dots, Y_m)^{\top}$ denote the vector of response
variables. Let $x_{i}$ and $z_{i}$ be the $p \times 1$ and
$q \times 1$ vectors of fixed and random effects covariates associated
with the $i$th response, respectively, $ i = 1, \dots, m$. Let
$\beta \in \mathbb{R}^{p}$ be the regression coefficients vector and
$u \in \mathbb{R}^{q}$ be the random effects vector. A GLMM can be
built with a link function $g$ that connects $\mu_i$, the
(conditional) expectation of $Y_i$, with the covariates $x_i, z_i$
satisfying $g(\mu_i) = x_i^{\top}\beta + z_i^{\top}u$, and assuming
that conditional on the random effects $u$, the responses $Y_i$'s are
independent with
\begin{equation}
  \label{eq:disy}
 Y_{i} \mid \beta,u \overset{ind}{\sim} f(y_{i}|\beta,u) \equiv \exp\Big\{\frac{y_i \xi_i - b(\xi_i)}{a_i(\iota)} + c_i(y_i, \iota)\Big\}  \quad  \text{for} \quad i = 1,...,m,
\end{equation}
where $\iota$ is the dispersion parameter and
$a_i(\cdot), b(\cdot), c_i(\cdot, \cdot)$ are known functions. To
simplify the presentation, here we assume that $\iota$ is known. The
quantity $\xi_i$ is associated with the conditional mean $\mu_i$ and
hence with $\beta, u$.  The description of a GLMM is completed by
specifying the distribution of $u$ and the exponential family pdf
\eqref{eq:disy} with the forms of the functions $a_i(\cdot), b(\cdot)$
and $c_i(\cdot, \cdot)$.  Assume there are $r$ random effects
$u_{1}^\top,u_{2}^\top,...,u_{r}^\top$, where $u_{j}$ is a
$q_{j} \times 1$ vector with $q_{j} > 0$, with
$q_{1} + q_{2} + ...+q_{r} = q$ and
$u = (u_{1}^\top,\dots,u_{r}^\top)^\top$. A common assumption is that
$u_{j} \overset{ind}{\sim} N(0, \Lambda_j \otimes R_j)$ where the
low-dimensional covariance matrix $\Lambda_j$ is unknown and need to be
estimated, and the structured matrix $R_j$ is usually known. Here,
$\otimes$ indicates the Kronecker product. Let
$y = (y_{1},y_{2},...,y_{m})$ denote the vector of observed
responses. For specific examples of \eqref{eq:disy} and the link
function $g(\cdot)$, we consider the three most popular GLMMs, namely
the logistic GLMM, the probit GLMM and the Poisson GLMM with the log
link.
\begin{example}[Logistic GLMM]
  \label{exa:logi}
  For the logistic GLMM, \eqref{eq:disy} is the binomial pmf given by
  \begin{equation}
    \label{eq:binlik}
  {\ell_i \choose
    y_i}(\mu_i/\ell_i)^{y_i}(1-\mu_i/\ell_i)^{\ell_i-y_i},
  y_i=0,1,\dots,\ell_i,  
  \end{equation}
  where $\ell_i$ is a positive integer indicating the number of
  trials. The logistic GLMM uses the logit link
  $g(\mu_i)= \log(\mu_i/[\ell_i-\mu_i])=x_i^{\top}\beta +
  z_i^{\top}u$.
\end{example}
\begin{example}[Probit GLMM]
   \label{exa:prob}
  In this case, \eqref{eq:disy} is also the binomial pmf \eqref{eq:binlik}
whereas, the probit link
$g(\mu_i)= \Phi^{-1}(\mu_i/\ell_i)=x_i^{\top}\beta + z_i^{\top}u$ is
used in the probit GLMM.  Here, $\Phi (\cdot)$ is the cdf of the standard normal distribution. 
\end{example}
\begin{example}[Poisson-log GLMM]
   \label{exa:pois}
 For the Poisson-log link model,
 \eqref{eq:disy} is the Poisson pmf given by
 \begin{equation}
   \label{eq:poislik}
   \exp(-\mu_i)\mu_i^{y_i}/y_i!, y_i=0,1,\dots,   
 \end{equation}
 with
$g(\mu_i) = \log(\mu_i) =x_i^{\top}\beta + z_i^{\top}u$.
 
\end{example}
 
Let
$\Lambda = (\Lambda_{1},\dots,\Lambda_{r})$. Then the likelihood function for $(\beta,\Lambda)$ is given by  
\begin{align}
\label{eq:likef}
L(\beta,\Lambda \mid y)= \int_{\mathbb{R}^{q}} \Bigg[\prod_{i=1}^{m} f(y_{i}|\beta,u)\Bigg] \phi_{q}(u; 0, G) du.
\end{align}
Here, $\phi_{q}(u;0,D)$ denotes the probability density function of
the $q$-dimensional normal distribution with mean vector $0$,
covariance matrix $D$, evaluated at $u$. Also,
$G =\oplus_{j=1}^{r} \Lambda_j \otimes R_j$, with $\oplus$ indicating
the direct sum. Outside the linear mixed model where \eqref{eq:disy}
is the normal density, $L(\beta,\Lambda \mid y)$ in \eqref{eq:likef}
nearly always involves intractable integrals and is not available in
closed-form.

There are two widely used Monte Carlo approaches for approximating the likelihood
function \eqref{eq:likef} and making inference on $(\beta, \Lambda)$, namely
the Monte Carlo EM algorithm \citep{boot:hobe:1999} and the Monte
Carlo maximum likelihood based on importance sampling
\citep{geye:thomp:1992, geye:1994a}. The EM is
an iterative method and each iteration of this algorithm consists of
an `E-step' and an `M-step'. The $(j+1)$st E-step entails the
calculation of
\begin{equation}
  \label{eq:mcem}
  E[\log f(y, u | \beta,\Lambda )|\beta^{(j)},\Lambda^{(j)}],
\end{equation}
where
\begin{equation}
  \label{eq:complik}
  f(y, u | \beta,\Lambda ) = \Bigg[\prod_{i=1}^{m} f(y_{i}|\beta,u)\Bigg] \phi_{q}(u;0,G)
\end{equation}
is the joint density of $(y, u)$ and $(\beta^{(j)}, \Lambda^{(j)})$ is the value of
$(\beta, \Lambda)$ from the $j$th iteration. The expectation in \eqref{eq:mcem} is with respect to
$f(u|y, \beta^{(j)}, \Lambda^{(j)})$, where $f(u|y, \beta, \Lambda)$ is the conditional density of $u$ given $y$
\begin{equation}
  \label{eq:postu}
  f(u|\beta, \Lambda, y) = \frac{f(y, u | \beta,\Lambda )}{L(\beta,\Lambda | y)},
\end{equation}
with $f(y, u | \beta,\Lambda )$ and $L(\beta,\Lambda \mid y)$ given in
\eqref{eq:complik} and \eqref{eq:likef}, respectively. Since
closed-form expressions of \eqref{eq:likef} and hence \eqref{eq:postu}
are not available, so are the means \eqref{eq:mcem} with respect to
the conditional density \eqref{eq:postu}. In the Monte
Carlo EM algorithm \citep{wei:tann:1990}, \eqref{eq:mcem} is
approximated by a Monte Carlo estimate which is then maximized in the
M-step. Indeed, if $\{u^{(n,j)},n= 1, \dots, N\}$ are samples obtained by running a Markov chain with
invariant density $f(u|\beta^{(j)}, \Lambda^{(j)},y)$, then a Monte
Carlo estimate of \eqref{eq:mcem} is
$\sum_{n=1}^N \log [f(y, u^{(n,j)} | \beta,\Lambda )]/N$ \citep{mccu:1994}.

In the Monte Carlo maximum likelihood \citep{geye:thomp:1992,
  geye:1994a}, the likelihood function \eqref{eq:likef} is expressed as
\begin{align}
  \label{eq:mcml}
  L(\beta,\Lambda | y) &= \int_{\mathbb{R}^{q}} f(y, u | \beta, \Lambda) du \nonumber \\ &= \int_{\mathbb{R}^{q}} \frac{f(y, u | \beta, \Lambda)}{f(y, u | \beta^{(0)}, \Lambda^{(0)})} f(y, u | \beta^{(0)}, \Lambda^{(0)}) du \propto E\bigg[\frac{f(y, u | \beta, \Lambda)}{f(y, u | \beta^{(0)}, \Lambda^{(0)})}\bigg],
\end{align}
where the expectation is with respect to the density \eqref{eq:postu}
at some value $(\beta^{(0)}, \Lambda^{(0)})$. Since the expectation in
\eqref{eq:mcml} can be approximated by
$(1/N)\sum_{n=1}^N f(y, u^{(n,0)} | \beta,\Lambda )/f(y, u^{(n,0)} | \beta^{(0)},
\Lambda^{(0)})$, maximum likelihood estimates of $(\beta, \Lambda)$ are calculated by
maximizing this Monte Carlo approximation. \cite{geye:1994a}
recommends making a few pilot iterations to find an appropriate value
for $(\beta^{(0)}, \Lambda^{(0)})$.  The importance sampling technique has been
successfully used for analyzing spatial generalized linear mixed models (SGLMMs)
which are GLMMs with the random effects $u$ being derived from a spatial process
\citep[see e.g.][]{chri:2004, roy:evan:zhu:2016, roy:tan:fleg:2018,
  evan:roy:2019}.

\section{Conditional simulation for GLMMs}
\label{sec:condsim}
Both the Monte Carlo EM and the Monte Carlo maximum likelihood methods
for making inference on $(\beta, \Lambda)$ requires effective methods for
sampling from the density $f(u|y, \beta, \Lambda)$ given in
\eqref{eq:postu}. To simplify notations, we use $f(u|y)$ to denote
this conditional density. In the context of some numerical examples involving
SGLMMs, \cite{roy:zhan:2021} observe poor performance of random walk
Metropolis compared to some other Metropolis-Hastings algorithms for sampling from \eqref{eq:postu}.
In this section, we present different variants of the MALA and HMC
methods for exploring \eqref{eq:postu}. We also describe some data
augmentation algorithms for the two particular GLMMs, namely the logistic and the probit
GLMMs.

\subsection{MALA for GLMMs}
\label{sec:mala}
MCMC methods are based on discrete time Markov chains. For example, as
mentioned in Section~\ref{sec:lik}, both Monte Carlo EM and Monte
Carlo maximum likelihood methods require Markov chains
$\{u^{(n)}\}_{n \ge 1}$ with appropriate stationary
densities. However, often, there are great benefits to first
considering an appropriate continuous time stochastic process that
possesses desirable properties. In particular, one can specify these
continuous time processes by some differential equations as
illustrated in this Section as well as in Section~\ref{sec:hmc}. For
example, MALA is a discrete time Markov chain based on the Langevin
diffusion $u_t$ defined as
\begin{equation}
  \label{eq:lang}
    du_t=(1/2)\nabla\log f(u_t|y)dt+ds_t,  
\end{equation}
where $s_t$ is the $q-$dimensional standard Brownian motion. It is
known that $f(u|y)$ is stationary for $u_t$ given in
\eqref{eq:lang}. On the other hand, discretizations of
\eqref{eq:lang}, say, by using the Euler-Maruyama method may fail to
maintain the stationarity with respect to $f(u|y)$. MALA is an MH chain
$\{u^{(n)}\}_{n \ge 1}$ where, in each iteration, the proposal $u'$ is
drawn following a simple discretization of \eqref{eq:lang} given by
\begin{equation}
  \label{eq:maladis}
    u' = u^{(n-1)}+ \epsilon \nabla\log f(u^{(n-1)}|y)/2 + \sqrt{\epsilon}
  \upsilon^{(n)}
\end{equation}
for a chosen step-size $\epsilon$ with
$\upsilon^{(n)} \stackrel{iid}{\sim} N(0, I_q)$. The proposal
$u'$ is accepted with probability
\begin{equation}
  \label{eq:accpro}
      \alpha(u^{(n-1)}, u')=1\wedge\frac{f(u'|y) k(u',u^{(n-1)})}{f(u^{(n-1)}|y) k(u^{(n-1)},u')},
\end{equation}
where the proposal density $k(u,u')$ is the
$N(u+\epsilon\nabla\log f(u|y)/2, \epsilon I_q)$ density evaluated at
$u'$.  Since the mean of the proposal density of the MALA is governed
by the gradient of log of the target distribution, it is likely to
make moves in the directions in which $f$ is increasing. That way, the
chain is encouraged to move towards the nearest mode of $f(u|y)$ and stay near the high mass regions of the target
density.
\begin{algorithm}[H]
\caption{The $n$th iteration for the MALA}
\begin{algorithmic}[1]
  \label{alg:mala}
  \STATE Given $u^{(n-1)}$ draw
  $u' \sim N(u^{(n-1)}+ \epsilon \nabla\log f(u^{(n-1)}|y)/2, \epsilon I_q)$.
  \STATE Draw $\delta \sim$ Uniform $(0, 1)$. If
  $\delta < \alpha(u^{(n-1)}, u')$ then set $u^{(n)} \leftarrow u'$, else
  set $u^{(n)} \leftarrow u^{(n-1)}$. Here, $\alpha(\cdot, \cdot)$ is as defined in \eqref{eq:accpro}.
\end{algorithmic}
\end{algorithm}
The MALA chain $\{u^{(n)}\}_{n \ge 1}$ can be used to approximate
\eqref{eq:mcem} or \eqref{eq:mcml}. In particular, if
$f(u|\beta^{(j)}, \Lambda^{(j)},y)$ is used for the stationary density
$f(u|y)$ in \eqref{eq:maladis} then the corresponding MALA chain can
be used for the $j$th E-step in the EM algorithm and when $f(u|y)$ in
\eqref{eq:maladis} is replaced with
$f(u|\beta^{(0)}, \Lambda^{(0)},y)$, it results in a MALA chain that
can be used for estimating \eqref{eq:mcml}. In practice, the step-size
$\epsilon$ is chosen as $O(q^{-1/3})$ obtaining an acceptance rate of
between 40\% and 70\% \citep{robe:rose:1998}. For implementing MALA we
need the derivatives of $\log f (u | y)$. Here, we derive
$\nabla\log f(u|y)$ for the three popular GLMMs mentioned in
Section~\ref{sec:lik}. Let $X$ and $Z$ be the $m \times p$ and
$m \times q$ known design matrices with $i$th row being $x_{i}^\top$
and $ z_{i}^\top, i=1,\dots,m$ respectively. Let
$\gamma_i = x_{i}^\top \beta + z_{i}^\top u, i=1,\dots,m$ and
$\gamma = X \beta + Z u$.

\begin{example}[continues=exa:logi]
For the binomial-logit link model, the
$\log f(u|y)$ (up to an additive constant) is
\begin{equation}
  \label{eq:lbinlogi}
    -\big[u^{\top}G^{-1}u +\log|G|\big]/2 +\sum_{i=1}^m \big[ y_i \gamma_i-\ell_i\log(1+\exp(\gamma_i))\big].
\end{equation}
% where $D$ is the matrix of covariates and $\Sigma \equiv \Sigma_\theta$ is the covariance matrix.
Letting $\xi$ be the $m \times 1$ vector with $i$th element
$\ell_i\exp(\gamma_i)/(1+\exp(\gamma_i)), i=1,\dots,m$, we have
\begin{equation*}
  \label{eq:binlder}
    \nabla\log f(u|y)= -G^{-1}u + Z^{\top} y - Z^{\top} \xi.
  \end{equation*}  
\end{example}
\begin{example}[continues=exa:prob]
  For the binomial-probit model, the $\log f(u|y)$ (up to an additive constant)
  is
  \begin{equation}
    \label{eq:lbinprob}
    -\big[u^{\top}G^{-1}u +\log|G|\big]/2 +\sum_{i=1}^m \big[ y_i \log(\Phi(\gamma_i)) + (\ell_i- y_i)\log(1 -\Phi(\gamma_i))\big].
\end{equation}
% where $D$ is the matrix of covariates and $\Sigma \equiv \Sigma_\theta$ is the covariance matrix.
Let $\tau_1$ and $ \tau_2$ be the $m \times 1$ vectors with $i$th
element $y_i\phi(\gamma_i)/\Phi(\gamma_i)$ and
$ [\ell_i -y_i]\phi(\gamma_i)/[1-\Phi(\gamma_i)],$ respectively,
$i=1,\dots,m$. Here, $\phi(\cdot) \equiv \phi_1(\cdot)$ is the
standard normal pdf. For the probit GLMM we have
\begin{equation*}
    \nabla\log f(u|y)= -G^{-1}u + Z^{\top} \tau_1 - Z^{\top} \tau_2.
\end{equation*}
\end{example}
\begin{example}[continues=exa:pois]
For the Poisson-log
GLMM, we derive $\log f (u |y)$ which (up to an additive constant) is 
\begin{equation}
   \label{eq:lpoislog}
\log f(u|y)= -\big[u^{\top}G^{-1}u +\log|G|\big]/2 + \sum_{i=1}^m (y_i\gamma_i-\exp\{\gamma_i\}),
\end{equation}
implying
\begin{equation*}
  \label{eq:poisder}
    \nabla\log f(u|y)=-G^{-1}u + Z^{\top} y - Z^{\top} \exp(\gamma).
\end{equation*}  
\end{example}
There are other variants of MALA, e.g., the pre-conditioned MALA
\citep{robe:stra:2007}, and the manifold MALA \citep{giro:cald:2011} proposed in the literature.
In the pre-conditioned MALA, the proposal density is
$N(u+ h M\nabla\log f(u|y)/2, hM)$ for some positive definite matrix
$M$. For conditional simulation in SGLMMs, \cite{roy:zhan:2021} observe
that the pre-conditioned MALA with an appropriately chosen $M$, and the
manifold MALA can have superior performance over the standard MALA.

%$\ell=(\ell_1,\dots,\ell_m)$

\subsection{HMC for GLMMs}
\label{sec:hmc}
In the HMC algorithm \citep{duan:kenn:pend:rowe:1987, neal:2011}, an
auxiliary variable $\rho \sim f(\rho) \equiv N(0, M)$ is introduced
for some $q \times q$ positive definite matrix $M$. The HMC chain
$\{u^{(n)}, \rho^{(n)}\}_{n \ge 1}$ alternates between draws from
$f(\rho^{(n)} | u^{(n-1)}) \equiv f(\rho^{(n)}) \equiv N(0, M)$ and
$f(u^{(n)} | \rho^{(n)}, u^{(n-1)})$. We now describe how a draw from
$f(u^{(n)} | \rho^{(n)}, u^{(n-1)})$ is made using ideas from
classical mechanics. The negative of logarithm of the joint density of
$u$ and $\rho$ given by
 \[
   H(u, \rho) = - \log f(u|y) + ( \log( (2\pi)^q |M|) + \rho^{\top} M^{-1}
   \rho)/2
 \]
 is a Hamiltonian function of the position ($u$) and the momentum
 ($\rho$). Intuitively, the Hamiltonian $H(u, \rho)$ measures the total
 energy of a physical system and it consists of the potential energy
 $- \log f(u|y)$ and the kinetic energy $\rho^{\top} M^{-1} \rho/2$. This is the
 reason $M$ is referred to as the mass matrix. The Hamiltonian
 equations are given by the following first order differential equations
   \begin{equation}
     \label{eq:hameqn}
     \frac{du}{dt} = \frac{\partial H}{\partial \rho} = M^{-1} \rho \;\;\mbox{and}\;\;  \frac{d \rho}{dt} = -\frac{\partial H}{\partial u } = \nabla\log f (u|y).
   \end{equation}
Solution of the Hamiltonian
 equations \eqref{eq:hameqn} results in the Hamiltonian flow from an initial $(u_0, \rho_0)$ to
   $(u_t, \rho_t)$. Here, $du/dt$ and $d\rho/dt$ denote the derivatives of
   $u$ and $\rho$ with respect to the (fictitious) continuous time $t$.
   To garner intuition behind \eqref{eq:hameqn} the following analogy
   in $\mathbb{R}^2$ is useful \citep{neal:2011}. Imagine a sledge
   sliding over a friction-less surface of varying height proportional
   to $1/f(u|y)$. The potential energy is based on the height of
   the surface at the current position, $u$, whereas the kinetic
   energy is determined by the sledge's momentum, $\rho$, and its mass,
   $M$. In a flat surface, that is, when
   $\nabla \log f(u|y) =0, \forall u$, the sledge moves at a constant
   velocity. On the other hand, when slope is positive
   ($\nabla \log f(u|y) <0$), the kinetic energy decreases as the
   potential energy increases until it vanishes ($\rho =0 $). The sledge
   then slides back down the hill increasing its kinetic energy and
   decreasing the potential energy.
   
Over any interval, the Hamiltonian dynamics \eqref{eq:hameqn} defines
the Hamiltonian flow $(u_0, \rho_0) \rightarrow (u_t, \rho_t)$ that
satisfies three important properties, namely, (i) it is energy
preserving, that is, $H(u_t, \rho_t) =H(u_0, \rho_0)$, (ii) it is volume preserving, that
is, $du_td\rho_t = du_0d\rho_0$ and (iii) it is time reversible, which implies
if, $(u_0, \rho_0) \sim \nu$ then $(u_t, \rho_t) \sim \nu$. Since \eqref{eq:hameqn} can not be solved analytically for
   practical examples, the St\"{o}rmer-Verlet or the leapfrog method is a
   standard approach for approximating the solutions to
   \eqref{eq:hameqn} \citep{duan:kenn:pend:rowe:1987}. In particular,
   this method uses a discrete step-size $\epsilon$ to make a move,
   according to
   \begin{align}
     \label{eq:leap}
     \rho_{t+\epsilon/2} &= \rho_t + (\epsilon/2)  \nabla\log f(u_t|y) \nonumber\\
     u_{t+\epsilon} &= u_t + \epsilon M^{-1}\rho_{t+\epsilon/2}\nonumber\\
      \rho_{t+\epsilon} &= \rho_{t+\epsilon/2} + (\epsilon/2)  \nabla\log f(u_{t+\epsilon}|y).
   \end{align}
   In HMC, in order to draw from $f(u^{(n)} | \rho^{(n)}, u^{(n-1)})$,
   starting from $(u^{(n-1)}, \rho^{(n)})$, the above set of
   deterministic steps \eqref{eq:leap} (referred to as Leapfrog
   $(\cdot, \cdot, \epsilon, M)$) is repeated $L$ times to generate a proposal
   $(u', \rho')$ which is then accepted/rejected with an MH step. The
   Leapfrog method preserves the volume exactly and it is also
   reversible by simply negating $\rho$ \cite[see][for
   details]{neal:2011}.
   
\begin{algorithm}[H]
\caption{The $n$th iteration for the HMC}
\begin{algorithmic}[1]
  \label{alg:hmc}
  \STATE Draw $\rho^{(n)} \sim N(0, M)$.
  \STATE Set $u'  \leftarrow u^{(n-1)}$ $\rho'  \leftarrow \rho^{(n)}$.
  \STATE For $i= 1,\dots,L$ do $(u', \rho') \leftarrow $ Leapfrog $(u', \rho', \epsilon, M)$.
%  \STATE Set $u^{*} \leftarrow u'$ and  $\rho^{*} \leftarrow -\rho'$.
  \STATE $\alpha \leftarrow $ min $(1, \exp\{-H(u', \rho') + H(u^{(n-1)}, \rho^{(n)})\})$
  
  \STATE Draw $\delta \sim$ Uniform $(0, 1)$. If $\delta < \alpha$
  then set $u^{(n)} \leftarrow u'$, $\rho^{(n)} \leftarrow -\rho'$, else set $u^{(n)} \leftarrow u^{(n-1)}$.
\end{algorithmic}
\end{algorithm}
Note that, if we could simulate the Hamiltonian dynamics
\eqref{eq:hameqn} exactly, by the energy preserving property (i), energy would be
preserved exactly, and the MH acceptance probability would always be min
$\{1, \exp(0)\} =1.$ Since we use the leapfrog integrator, that
approximately simulates the dynamics, if the approximation is good,
then $H(u', \rho') - H(u^{(n-1)}, \rho^{(n)})$ would be small, and the
acceptance rate will be high. Indeed, for HMC, $\alpha$ still tends
to be high even for proposals that are far from the current state, reducing the
random walk behavior of some other MH algorithms. The marginal chain $\{u^{(n)}\}_{n \ge 1}$ of
the HMC chain $\{u^{(n)}, \rho^{(n)}\}_{n \ge 1}$ can be used to approximate the expectations in \eqref{eq:mcem} or
\eqref{eq:mcml}.

The choices of $\epsilon, L$ and the mass matrix $M$ should be such
that the resulting algorithm mixes well (that is, the sampled
distribution is `close' to the target distribution), leads to suitable
acceptance rates and lower Monte Carlo errors. Often, in practice, the
mass matrix $M$ is chosen to be the identity matrix and $\epsilon, L$
are adjusted to achieve around 70\% acceptance rates. The No-U-Turn
sampler (NUTS) \citep{hoff:gelm:2014} is an extension of HMC that
eliminates the need of manual tuning of $L$. \cite{hoff:gelm:2014}
also propose a method for dynamically adapting the $\epsilon$
parameter on the fly. NUTS is employed in the programming language
Stan \citep{carp:gelm:2017}. \cite{giro:cald:2011} propose
the Riemannian manifold HMC algorithm that uses a position-dependent
$M$ that changes in every iteration, eliminating the need for manually
tuning the mass matrix. A comparison of performance of different HMC
algorithms in the context of analyzing GLMMs can be found in
\cite{zhan:2022}.

\subsection{Data augmentation for GLMMs}
\label{sec:da}
Data augmentation (DA) is an MCMC algorithm that has been widely used
for analyzing Bayesian probit and logistic GLMs \citep[see
e.g.][]{albe:chib:1993, pols:scot:wind:2013}. Recently, DA algorithms
have been developed and studied for Bayesian GLMMs
\citep{wang:roy:2018, pols:scot:wind:2013, wang:roy:2018a,
  rao:roy:2021}. In this section, we propose DA algorithms for
simulating from \eqref{eq:postu} corresponding to the probit and
logistic mixed models. For constructing a valid and efficient DA algorithm
\citep{tann:wong:1987} for $f(u|\beta, \Lambda, y)$ in \eqref{eq:postu} we
need to construct a joint density $f(u, d|\beta, \Lambda, y)$ with augmented
variables $d$ satisfying the following two properties
\begin{enumerate}
\item[(i)] the
$u-$marginal of the joint density $f(u, d|\beta, \Lambda, y)$ is the target density
\eqref{eq:postu} and
\item[(ii)] sampling from the two corresponding conditional
densities $f_{u|d}$ and $f_{d|u}$ is straightforward. 
\end{enumerate}
Each iteration of the DA algorithm consists of two steps --- a draw
from $f_{d|u}$ followed by a draw from $f_{u|d}$. Thus, the DA Markov
chain $\{u^{(n)}, d^{(n)}\}_{n \ge 1}$ is a two-variable Gibbs
sampler. The DA algorithm, like its deterministic counterpart the EM
algorithm, is widely used. In Sections~\ref{sec:acda} and
\ref{sec:pgda}, we provide appropriate DA algorithms for the probit
and logistic mixed models, respectively.

\subsubsection{Data augmentation for probit mixed models}
\label{sec:acda}
In this Section, we consider probit GLMM for binary data, that is,
$\ell_i =1$ for $i=1,\dots,m$ in Example~\ref{exa:prob}. Thus,
$(Y_1, Y_2, \dots, Y_m)$ are independent Bernoulli random variables
with $P(Y_i =1) = \Phi(x_i^{\top}\beta + z_i^{\top}u)$. Following
\cite{albe:chib:1993}, let $v_i \in \mathbb{R}$ be the continuous latent variable
corresponding to binary observation $Y_i$, such that
$Y_i = I(v_i >0)$, where
$v_i | \beta, u \overset{\text{ind}} \sim N( \gamma_i$, 1) for
$i =1,\dots, m$. Then
\begin{equation}
  \label{eq:aclate}
P(Y_i=1) = P(v_i >0) =\Phi(\gamma_i).  
\end{equation}
 Let
$v = (v_1,\dots, v_m)^{\top}$, then
$v| \beta, u \sim N(X\beta + Zu,I_m)$. Using the latent variables $v$, we introduce the joint density
\begin{align}
\label{eq:probjt}
  f(u, v|\beta, \Lambda, y) = \frac{1}{L(\beta, \Lambda|y)}\left[ \prod_{i=1}^m \phi(v_i; \gamma_i, 1) \left[1_{\text{(0,\ensuremath{\infty})}}\left(v_{i}\right)\right]^{y_{i}}\left[1_{\left(-\infty,0\right]}\left(v_{i}\right)\right]^{1-y_{i}}\right]\times \phi_q(u; 0, G).
\end{align}
From \eqref{eq:aclate} it follows that
\begin{equation}
  \label{eq:acdacond}
    \int_{\mathbb{R}^m} f(u, v|\beta, \Lambda, y) dv = f(u |\beta, \Lambda, y),
  \end{equation}
  where $f(u |\beta, \Lambda, y)$, given in \eqref{eq:postu}, is the target density.
Thus, the condition (i) of DA construction mentioned before holds. From \eqref{eq:probjt}, it follows that
\begin{equation}
\label{eq:problatcond}
v_i| u, \beta, \Lambda, y \overset{\text{ind}}\sim \text{TN}(\gamma_i,1,y_i), \, i=1,\dots, m,
\end{equation}
where $\text{TN}(\mu, \sigma^2, e)$ denotes the distribution of
the normal random variable with mean $\mu$ and variance $\sigma^2$,
that is truncated to have only positive values if $e = 1$,
and, it has only negative values if $e = 0$.

From \eqref{eq:probjt} it follows that the conditional density of $u$ given $\beta, \Lambda, v, y$ is
\begin{align*}
f(u \mid \beta, \Lambda, v, y) &\propto \left[\prod_{i=1}^{m} \exp\Big\{ -\frac{1}{2}\Big[ (z_{i}^\top u)^{2} - 2 (z_{i}^\top u)(v_i -x_{i}^\top \beta)\Big] \Big\}\right] \exp\Big[ -\frac{1}{2}u^\top G^{-1}u \Big]\\
&= \exp \Big[  - \frac{1}{2} \Big\{ u^\top (Z^\top Z + G^{-1}) u - 2 u^\top Z^\top (v - X \beta) \Big\}\Big].
\end{align*}
Thus, the conditional distribution of $u$ is
\begin{equation}
  \label{eq:probucond}
  u|v, \beta, \Lambda, y \sim N_{q}\left( (Z^{\top}Z + G^{-1})^{-1} Z^{\top} (v - X\beta), (Z^{\top}Z + G^{-1})^{-1}\right).
\end{equation}
Thus, every iteration of the proposed DA algorithm for
\eqref{eq:probjt} consists of making the draws of $v$ and $u$
from \eqref{eq:problatcond} and \eqref{eq:probucond}, respectively.
\begin{algorithm}[H]
  \caption{The $n$th iteration for the DA algorithm}
  \label{alg:probda}
\begin{algorithmic}[1]
  \STATE Given $u^{(n-1)}$, draw $v_i^{(n)} \overset{\text{ind}}\sim \text{TN}(x_i^{\top}\beta+z_i^{\top} u^{(n-1)},1,y_i)$ for  $i=1,\dots, m$.
  %draw $v^{(n)} $ from \eqref{eq:problatcond} with $u = u^{(n-1)}$.
%\STATE[1]
%\hspace{.18in}
\STATE Draw $u^{(n)} \sim N_{q}\left( (Z^{\top}Z + G^{-1})^{-1} Z^{\top} (v^{(n)} - X\beta), (Z^{\top}Z + G^{-1})^{-1}\right)$.
\end{algorithmic}
\end{algorithm}

The conditional distribution of $u$ in the above DA algorithm and several other
conditional distributions appearing in this article are
normal distributions of the form $N(S^{-1} t, S^{-1})$ for some
positive definite matrix $S$ and a vector $t$. A naive method of drawing from $N(S^{-1} t, S^{-1})$ is inefficient
when the dimension of $S$ is large as it involves calculating inverse
of the matrix $S$. \cite{rao:roy:2021} advocate using the following method of drawing
from $N(S^{-1} t, S^{-1})$ that does not require computing
$S^{-1}$. 
  \begin{algorithm}[H]
\caption{An algorithm for drawing from $N(S^{-1} t, S^{-1})$}
\begin{algorithmic}[1]
\STATE Let $S = LL^\top$ be the Cholesky decomposition of $S$.
\STATE Solve $L w = t$.
\STATE Draw  $z \sim N(0, \I_{q})$ where $q$ is the dimension of $S$.
\STATE Solve $L^\top x = w + z$. Then $x \sim N(S^{-1} t, S^{-1})$. 
\end{algorithmic}
\end{algorithm}

DA algorithms, although our popular, they often suffer from slow
convergence and high autocorrelations. \cite{liu:wu:1999} proposed the
parameter expansion for data augmentation (PX-DA) algorithms for
speeding up the convergence of DA algorithms. More recently,
\cite{hobe:marc:2008} compared the performance of PX-DA algorithms
based on a Haar measure (called the Haar PX-DA algorithms), the PX-DA
algorithms based on a probability measure and the DA
algorithms. \cite{hobe:marc:2008} showed that, under some mild
conditions, the Haar PX-DA algorithms are better than the PX-DA and the DA
algorithms in terms of different ordering. In PX-DA, an
extra step is added (sandwiched) between the two steps of the original
DA algorithm. In order to construct this extra step, we derive the
marginal density of $v$ from the joint density \eqref{eq:probjt} as
\begin{eqnarray}
\label{eq:margthet}
  f\left(v |y \right) & = & \int_{\mathbb{R}^{q}}f(u, v | y) du\\
                                           & \propto & \prod_{i=1}^{m}\left[1_{\text{(0,\ensuremath{\infty})}}\left(v_{i}\right)\right]^{y_{i}}\left[1_{\left(-\infty,0\right]}\left(v_{i}\right)\right]^{1-y_{i}} \exp\left\{ -\frac{1}{2}\left[v^{\top}Z_1v-2v^{\top}Z_1X\beta\right]\right\}, \nonumber
\end{eqnarray}
where
\[
Z_1 = \left[I_m-Z\left(Z^\top Z + G^{-1}\right)^{-1}Z^{\top}\right].
\]
%Let $\Psi^*$ be the
%corresponding Markov chain based on the Haar PX-DA algorithm,
 Let $\mathcal{V}$ denote the subset of $\mathbb{R}^m$ where $v$
lives, that is, $\mathcal{V}$ is the Cartesian product of $m$ half
(positive or negative) lines, where the $i$th component is
$(0, \infty)$ (if $y_i =1$) or $(-\infty, 0]$ (if $y_i =0$).  Let $\psi$
be the unimodular multiplicative group on $\mathbb{R}_+$ with Haar
measure $\nu(dh) = dh/h$, where $dh$ is Lebesgue measure on
$\mathbb{R}_+$.  For constructing an efficient extra step, as in
\cite{roy:hobe:2007}, we let the group $\psi$ act on
$\mathcal{V}$ through a group action
$T(v) = hv = (hv_1, hv_2, \dots,
hv_m)^{\top}$. With the group action defined this way, it can be
shown that the Lebesgue measure on $\mathcal{V}$
is relatively left invariant with the multiplier $\chi (h) = h^m$
\citep{roy:2014, hobe:marc:2008}. Following
\cite{hobe:marc:2008}, consider a probability density function
$\omega(h)$ on $\psi$ where
\begin{eqnarray}
\label{eq:gprob}
\omega\left(h\right) dh &\propto&  \nonumber f\left(h v| y\right) \chi \left( h \right) \nu(dh)\\
& \propto & h^{m-1}\exp\left\{ -\frac{1}{2}\left[h^{2}v^{\top} Z_1 v -  2h v^{\top} Z_1 X\beta \right]\right\} dh.
\end{eqnarray}
Since $Z_1$ is a positive definite matrix, given $v$,
$\omega\left(h\right)$ is a valid density. From \cite{hobe:marc:2008},
it follows that the transition $v \rightarrow v' \equiv T(v) = hv$
where $h \sim \omega(h)$, is reversible with respect to $f(v| y)$
defined in \eqref{eq:margthet}. As mentioned in \cite{roy:2014},
intuitively, the extra step \eqref{eq:gprob} reduces the correlation
between $u^{(n-1)}$ and $u^{(n)}$ and thus improves the mixing of the DA
algorithm. Below are the three steps involved in every iterations of
the proposed Haar PX-DA algorithm.

\begin{algorithm}[H]
\caption{The $n$th iteration for the Haar PX-DA algorithm}
\label{alg:probpxda}
\begin{algorithmic}[1]
  \STATE  Draw $v_i^{(n)} \overset{\text{ind}}\sim \text{TN}(x_i^{\top}\beta+z_i^{\top} u^{(n-1)},1,y_i)$ for  $i=1,\dots, m$.
%\STATE[1]
%\hspace{.18in}
  
\STATE Draw $h$ from \eqref{eq:gprob}.

\STATE Calculate
  $v_i^{\prime} = hv_i$ for $i=1,\dots,m$, and draw $u^{(n)}$ from \eqref{eq:probucond} conditional on
  $v^{\prime} = (v_1^{\prime},\dots, v_m^{\prime})^{\top}$, that is, draw
\[u^{(n)} \sim N_{q}\left( (Z^{\top}Z + G^{-1})^{-1} Z^{\top} (v' - X\beta), (Z^{\top}Z + G^{-1})^{-1}\right).\]
 \end{algorithmic}
\end{algorithm}

Since $\omega(h)$ in \eqref{eq:g} is log-concave, adaptive rejection
sampling algorithm \citep{gilk:wild:1992} can be used to efficiently
sample from $\omega(h)$. The only difference between the Haar PX-DA
algorithm (Algorithm~\ref{alg:probpxda}) and the DA algorithm
(Algorithm~\ref{alg:probda}) is a single draw from the univariate
density $\omega(h)$, which is easy to sample from. Thus, the
computational burden, per iteration, for the Haar PX-DA algorithm is
similar to that of the DA algorithm.

\subsubsection{ Data augmentation for logistic mixed models}
\label{sec:pgda}
Since the
highly cited paper of \cite{albe:chib:1993} for probit GLMs, there have been
several attempts to construct such a DA sampler for the logistic
model. Recently, \cite{pols:scot:wind:2013}
 have proposed an efficient DA Gibbs
sampler for Bayesian logistic models with P\'{o}lya-Gamma (PG) latent
variables. A random variable $\varphi$ has PG distribution with
parameters $a, b$, that is, $\varphi \sim $PG$(a,b)$, if
$\varphi \overset{d}{=} (1/(2\pi^2)) \sum_{i =1}^{\infty} \varphi_{i} /
[(i-1/2)^2 +b^2/(4\pi^2)]$, where $\varphi_{i} \overset{iid}{\sim}$
Gamma$(a,1), a>0, b \in \mathbb{R}$. From \cite{wang:roy:2018b}, the pdf for PG$(a,b)$ is 
\begin{align*}
p(\varphi \mid a,b) = \bigg[\cosh \bigg(\frac{b}{2}\bigg)\bigg]^a\frac{2^{a - 1}}{\Gamma{(a)}} \sum_{j=0}^{\infty}(-1)^j \frac{\Gamma{(j+a)}}{\Gamma{(j+1)}}\frac{(2j+a)}{\sqrt{2\pi \varphi^{3}}} \exp\Big(-\frac{(2j+a)^{2}}{8\varphi} - \frac{\varphi b^{2}}{2}\Big),
\end{align*}
for $\varphi>0,$ where the hyperbolic cosine function $\cosh(t) = (e^{t} + e^{-t})/2$.

\pcite{pols:scot:wind:2013} DA technique can be extended to construct a
Gibbs sampler for logistic GLMMs (Example~\ref{exa:logi}).
Indeed, from \cite{pols:scot:wind:2013} we have
\begin{equation}
  \label{eq:pglat}
  \frac{[\exp(\gamma_i)]^{y_i}}{[1+\exp(\gamma_i)]^{\ell_i}} = 2^{-\ell_i} \exp(\kappa_i \gamma_i) \int_{0}^{\infty} \exp[-w_i\gamma_i^2/2] p(w_i) dw_i,
\end{equation}
where $\kappa_{i} = y_{i} -1/2, \, i = 1,...,m$ and $p(w_{i})$ is the pdf of PG$(\ell_i,0)$. %  given by, 
% \begin{equation}
% \label{eq:pgell}
% p(w_{i})= \frac{2^{\ell_i -1}}{\Gamma(\ell_i)}\sum_{j = 0}^{\infty} (-1)^{j} \frac{\Gamma(j+\ell_i)}{\Gamma(j+1)}\frac{(2j+\ell_i)}{\sqrt{2\pi w_{i}^{3}}} \exp\big[-\frac{(2j+\ell_i)^{2}}{8w_{i}}\big],\,w_{i} > 0.
% \end{equation}
Using PG latent variables
$w = (w_{1},w_{2},...,w_{m})$, we construct the
joint density
\begin{equation}
  \label{eq:logijt}
  f(u, w|\beta, \Lambda, y) \propto  \bigg[\prod_{i=1}^{m}  \exp\{\kappa_{i} \gamma_i -w_{i}\gamma_i^{2}/2\}p(w_{i}) \bigg] \phi_{q}(u;0,G).
\end{equation}
From \eqref{eq:pglat} it follows that the $u-$ marginal of
\eqref{eq:logijt} is the target density $f(u |\beta, \Lambda, y)$.
The conditional density for $w_{i}$ is
\begin{equation*}
  \label{eq:pomegai}
f(w_{i} \mid u, \beta, \Lambda,y)
\propto \exp(-w_{i}\gamma_i^{2}/2) p(w_{i}).
\end{equation*}
From \cite{rao:roy:2021} we then have
\begin{align}
\label{eq:distributionomega}
w_{i} \mid u, \beta, \Lambda ,y \overset{ind}\sim \mbox{PG} (\ell_i,\gamma_i), \,i = 1,\dots, m.
\end{align}
\cite{pols:scot:wind:2013} describe an efficient method for sampling
from the PG distribution. Also, from \eqref{eq:logijt}, as in \cite{rao:roy:2021},
the conditional density of $u$ given $\beta, \Lambda, w, y$ is
\begin{align*}
f(u \mid \beta, \Lambda, w, y) &\propto \prod_{i=1}^{m} \exp\Big\{ \kappa_{i} z_{i}^\top u - \frac{w_{i}}{2}\big[ (z_{i}^\top u)^{2} + 2 (z_{i}^\top u)(x_{i}^\top \beta)\big] \Big\} \exp\Big[ -\frac{1}{2}u^\top G^{-1}u \Big]\\
&= \exp \Big[  - \frac{1}{2} u^\top (Z^\top W Z + G^{-1}) u + u^\top (Z^\top \kappa - Z^\top W X \beta)\Big],
\end{align*}
where $W$ is the $m \times m$ diagonal matrix with $i^{th}$ diagonal element $w_{i}$ and $\kappa = (\kappa_{1},\dots,\kappa_{m})^\top$. 
Thus, the conditional distribution of $u$ is
\begin{equation}
  \label{eq:logiucond}
  u|w, \beta, \Lambda, y \sim N_{q}\left( (Z^{\top} W Z + G^{-1})^{-1} (Z^{\top}\kappa- Z^{\top} W X\beta), (Z^{\top} W Z + G^{-1})^{-1}\right).
\end{equation}
So, every iteration of the P\'{o}lya Gamma sampler for
\eqref{eq:logijt} consists of making the draws of $w$ and $u$
from \eqref{eq:distributionomega} and \eqref{eq:logiucond}, respectively.

\begin{algorithm}[H]
\caption{The $n$th iteration for the DA algorithm}
\begin{algorithmic}[1]
  \STATE Given $u^{(n-1)}$, draw $\omega_i^{(n)} \overset{ind}\sim \mbox{PG} (\ell_i,x_i^{\top}\beta + z_i^{\top}u^{(n-1)}), \,i = 1,\dots, m.$
  % from \eqref{eq:distributionomega} with $u = u^{(n-1)}$.
%\STATE[1]
%\hspace{.18in}
\STATE Draw $u^{(n)} \sim N_{q}\left( (Z^{\top} W Z + G^{-1})^{-1} (Z^{\top}\kappa- Z^{\top} W X\beta), (Z^{\top} W Z + G^{-1})^{-1}\right)$ with $w = w^{(n)}$.
\end{algorithmic}
\end{algorithm}

\section{MCMC for Bayesian GLMMs}
\label{sec:bayes}
Here, we consider MALA, HMC and DA algorithms for Bayesian GLMMs.  In the Bayesian framework,
one needs to specify the prior distributions of $\beta$ and $\Lambda$. We assume
the Gaussian prior for $\beta$ given by
\begin{align}
\label{eq:betaprior}
f(\beta) \propto \exp \Big[-\frac{1}{2}(\beta-\mu_{0})^\top Q(\beta-\mu_{0}) \Big],
\end{align}
where $\mu_{0} \in \mathbb{R}^{p}$ and $Q$ is a $p \times p$ positive definite matrix.

For simplifying the presentations, we assume that the structured
matrices $R_j$'s to be identity matrices and the covariance matrices
$\Lambda_j$'s correspond to scalar variances. Thus,
$\Lambda_j \otimes R_j= (1/\lambda_{j})\I_{q_{j}}$, where $\lambda_{j} > 0$, that is, the
components in $u_j$ are independent with a common variance $1/\lambda_j$. Let
$\lambda = (\lambda_{1},...,\lambda_{r})$. We assume that the prior for $\lambda_{j}$ is 
\begin{align}
\label{eq:tauprior}
f(\lambda_{j}) \propto \lambda_{j}^{a_{j}-1}e^{-b_{j}\lambda_{j}}, \;j = 1,\dots,r,
\end{align}
for $a_j >0, b_j >0, $ that is, apriori $\lambda_j \sim $ Gamma $(a_j, b_j),$ $ j = 1,\dots,r$.  Finally, we assume that $\beta$
and $\lambda$ are apriori independent and all $\lambda_{j}$s are also
apriori independent. Hence, the joint posterior density for
$(u, \beta,\lambda)$ is
\begin{align}
\label{eq:jointpd}
  f(u, \beta,\lambda | y) & \propto f(y, u| \beta, \lambda) f(\beta) \prod_{j=1}^r f(\lambda_j) \nonumber\\
 & \propto \Bigg[\prod_{i=1}^m f(y_i| \beta, u)\Bigg] f(\beta) \Bigg[\prod_{j=1}^{r}\lambda_{j}^{a_j -1 +q_{j}/2} \exp[-(b_j + u_j^{\top} u_j/2)\lambda_{j}]\Bigg],
\end{align}
where $f(y_i| \beta, u), f(y, u| \beta, \lambda),  f(\beta)$ and $f(\lambda_j)$ are given in
\eqref{eq:disy}, \eqref{eq:complik}, \eqref{eq:betaprior}, and \eqref{eq:tauprior},
respectively.

In \eqref{eq:betaprior} if $Q = 0$, then $\pi(\beta) \propto 1$, that
is, in that case, \eqref{eq:betaprior} becomes the improper uniform
prior on $\beta$. Similarly, the prior on $\lambda$ in
\eqref{eq:tauprior} will be improper if $a_{j}$ and/or $b_{j}$ takes
non-positive values. Several of the MCMC algorithms presented here are
also applicable to the situations when $\pi(\beta) \propto 1$ and/or
$\pi(\lambda)$ in \eqref{eq:tauprior} is improper. But, we do not pursue
the use of improper priors here. Interested readers may look at
\cite{wang:roy:2018} and \cite{rao:roy:2021}. On the other hand, if
improper priors are used, then the posterior density
\eqref{eq:jointpd} is not guaranteed to be proper. Hence, in such
cases, it is necessary to show that \eqref{eq:jointpd} is a proper pdf
before carrying out further inference. Also, it is known that the
usual (sample average) Monte Carlo estimators converge to zero with
probability one if the MCMC chain corresponds to an improper target
distribution \citep{athr:roy:2014b}. We now present various MCMC algorithms for exploring the
posterior density \eqref{eq:jointpd}. 

\subsection{MALA and HMC for Bayesian GLMMs}
\label{sec:bmala}
The logarithm of the posterior density \eqref{eq:jointpd}  (up to an additive constant)
is
\begin{align}
  \label{eq:logjt}
  \log f(u, \beta,\lambda|y) = \sum_{i=1}^m \big[(y_i \xi_i & - b(\xi_i))/a_i(\iota)\big] -(\beta-\mu_{0})^\top Q(\beta-\mu_{0})/2 \nonumber\\ & + \sum_{j=1}^r \big[(a_{j}-1 + q_j/2) \log \lambda_j - (b_{j}+ u_j^{\top} u_j/2)\lambda_{j} \big].
\end{align}
We can construct a MALA for \eqref{eq:jointpd} following
Algorithm~\ref{alg:mala} given in Section~\ref{sec:mala} using the
derivatives of $\log f(u, \beta, \log(\lambda)|y)$, but we propose a
different algorithm. From \eqref{eq:jointpd}, we know that conditional
on $(u, \beta, y)$,
\begin{equation}
  \label{eq:taucond}
  \lambda_{j} \overset{ind}\sim \mbox{Gamma}(a_{j} + q_{j}/2, b_{j} + u_{j}^\top u_{j}/2), \,j = 1,...,r.
\end{equation}
Denoting $\zeta = (u, \beta)$ and
$\zeta^{(n)} = (u^{(n)}, \beta^{(n)})$, we suggest running a MALA
within Gibbs chain $\{\zeta^{(n)},\lambda^{(n)}\}_{n \ge 1}$, where each iteration
alternates between a MALA step for $f(\zeta|\lambda, y)$ and a draw of
$\lambda$ from \eqref{eq:taucond}. Here, $f(\zeta|\lambda, y)$ is the
conditional density of $\zeta$ given by
\begin{align*}
  f(\zeta|\lambda, y) &= f(u, \beta | \lambda, y)\\ & \propto \Bigg[\prod_{i=1}^m f(y_i| \beta, u)\Bigg] \exp \Big[-\frac{1}{2}(\beta-\mu_{0})^\top Q(\beta-\mu_{0}) \Big] \exp\Bigg[-\sum_{j=1}^r \lambda_j u_j^{\top} u_j/2\Bigg].
\end{align*}
\begin{algorithm}[H]
\caption{The $n$th iteration for the MALA}
\begin{algorithmic}[1]
  \label{alg:bmala}
  \STATE Given $(\zeta^{(n-1)}, \lambda^{(n-1)})$ draw
  $\zeta' \sim N(\zeta^{(n-1)}+ \epsilon \nabla\log
  f(\zeta^{(n-1)}|\lambda^{(n-1)}, y)/2, \epsilon I_{p+q})$.  \STATE
  Draw $\delta \sim$ Uniform $(0, 1)$. If
  $\delta < \alpha(\zeta^{(n-1)}, \zeta')$ then set
  $\zeta^{(n)} \leftarrow \zeta'$, else set
  $\zeta^{(n)} \leftarrow \zeta^{(n-1)}$. Here, $\alpha(\cdot, \cdot)$
  is obtained from \eqref{eq:accpro} by replacing $f(u|y)$ with
  $f(\zeta|\lambda^{(n-1)}, y)$ and $k(u, u')$ with $k(\zeta, \zeta')$
  which is $N(\zeta+\epsilon\nabla\log f(\zeta|\lambda^{(n-1)},y)/2, \epsilon I_{p+q})$ density evaluated at $\zeta'$.
  \STATE Draw $\lambda_{j}^{(n)} \overset{ind}\sim$
  Gamma$(a_{j} + q_{j}/2, b_{j} + u_{j}^\top u_{j}/2), \,j = 1,...,r$
  with $u = u^{(n)}$.
\end{algorithmic}
\end{algorithm}

For the three GLMM examples, $\nabla_u \log f (\zeta| \lambda, y)$, the
derivatives of $\log f (\zeta| \lambda, y)$ with respect to $u$ are given
in Section~\ref{sec:mala} with $G= D(\lambda)^{-1}$ where
$D(\lambda) = \oplus_{j=1}^{r} \lambda_{j} \I_{q_{j}}$. Here, we derive
$\nabla_\beta \log f (\zeta| \lambda, y)$ for these three popular GLMMs.
\begin{example}[continues=exa:logi]
For the binomial-logit link model, from \eqref{eq:lbinlogi} and \eqref{eq:logjt} it follows that
\begin{equation*}
  \nabla_\beta \log f(\zeta|\lambda,y)= X^{\top} y - X^{\top} \xi - Q(\beta-\mu_{0}).
  %, \nabla_\lambda \log f(u, \beta,\lambda|y)= \sum_{j=1}^r \big[(a_{j}-1 + q_j/2)/\lambda_j - (b_{j}+ u_j^{\top} u_j/2)\lambda_{j} \big]..
  \end{equation*}  
\end{example}
\begin{example}[continues=exa:prob]
  For the binomial-probit model, from \eqref{eq:lbinprob} and \eqref{eq:logjt} it follows that
\begin{equation*}
    \nabla_\beta \log f(\zeta|\lambda,y) = X^{\top} \tau_1 - X^{\top} \tau_2 - Q(\beta-\mu_{0}).
\end{equation*}
\end{example}
\begin{example}[continues=exa:pois]
For the Poisson-log
GLMM, from \eqref{eq:lpoislog} and \eqref{eq:logjt} we have
\begin{equation*}
    \nabla_\beta \log f(\zeta|\lambda,y)=X^{\top} y - X^{\top} \exp(\gamma)- Q(\beta-\mu_{0}).
\end{equation*}  
\end{example}
Also, in this case, we propose a HMC within Gibbs chain
$\{\zeta^{(n)},\lambda^{(n)}\}_{n \ge 1}$, where each iteration
alternates between a HMC step for $f(\zeta|\lambda, y)$ as in
Algorithm~\ref{alg:hmc} for the Hamiltonian function
$H(\zeta, \rho) = - \log f(\zeta|\lambda,y) + [(p+q) \log(2\pi) + \log
(|M|) + \rho^{\top}M^{-1}\rho]/2 $ with $\rho \sim N(0, M)$ for
a $(p+q) \times (p+q)$ positive definite matrix $M$ and a draw of
$\lambda$ from \eqref{eq:taucond}.
% \subsection{HMC for Bayesian GLMMs}
% \label{sec:bhmc}

\subsection{Data augmentation for Bayesian GLMMs}
\label{sec:bda}
As mentioned in Section~\ref{sec:da}, for a successful DA for Bayesian
GLMMs, we need to construct a joint density $f(u, \beta,\lambda, d | y)$
with augmented variables $d$ whose $(u, \beta,\lambda)-$ marginal is the
density \eqref{eq:jointpd}. In this Section, we show that the
augmented variables derived in sections~\ref{sec:acda} and
\ref{sec:pgda} can be used for constructing DA for the Bayesian probit
and logistic mixed models, respectively.

\subsubsection{Data augmentation for Bayesian probit mixed models}
\label{sec:bacda}
As in Section~\ref{sec:acda}, we consider a vector of Bernoulli random
variables $(Y_1, Y_2, \dots, Y_m)$ and assume
$P(Y_i =1) = \Phi(\gamma_i), i=1,\dots,m$.  Using the latent variables $v$
introduced in Section~\ref{sec:acda}, \cite{wang:roy:2018} introduce
the joint density
\begin{align}
\label{eq:bprobjt}
f(u, \beta, \lambda, v|y) &\propto \left[ \prod_{i=1}^m \phi(v_i;x_i^{\top}\beta + z_i^\top u, 1) \left[1_{\text{(0,\ensuremath{\infty})}}\left(v_{i}\right)\right]^{y_{i}}\left[1_{\left(-\infty,0\right]}\left(v_{i}\right)\right]^{1-y_{i}}\right] \nonumber\\ & \times \phi_q(u; 0, D(\lambda)^{-1})f(\beta) \prod_{j=1}^r f(\lambda_j).
\end{align}
From \eqref{eq:acdacond} it follows that
\[
  \int_{\mathbb{R}^m} f(u, \beta, \lambda, v|y) dv = f(u, \beta, \lambda|y).
\]
Thus, $(u, \beta,\lambda)-$ marginal of
\eqref{eq:bprobjt} is the target density \eqref{eq:jointpd}.

From \eqref{eq:bprobjt}, the conditional density of $\beta$ given $u, \lambda, v, y$ is 
\begin{align*}
f(\beta \mid  u, \lambda, v, y) &\propto \left[\prod_{i=1}^{m} \exp\Big\{ -\frac{1}{2}\Big[ (x_{i}^\top \beta)^{2} - 2 (x_{i}^\top \beta)(v_i -z_{i}^\top u)\Big] \Big\}\right]\\ & \times \exp\Big[-\frac{1}{2}(\beta-\mu_{0})^\top Q(\beta-\mu_{0})\Big] \\
                                  &\propto \exp \Big[  - \frac{1}{2} \Big\{ \beta^\top (X^\top X + Q) \beta - 2 \beta^\top (X^\top v - X^\top Z u + Q\mu_0) \Big\}\Big].
\end{align*}
Thus, the conditional distribution of $\beta$ given $u, \lambda, v, y$ is 
\begin{align}
\label{eq:probbetcond}
\beta \mid u,\lambda, v, y \sim N((X^\top X + Q)^{-1} (X^\top v + Q \mu_{0} - X^\top Z u),(X^\top X+ Q)^{-1}).
\end{align}
The conditional densities of $v, u$ and $\lambda$ are given in
\eqref{eq:problatcond}, \eqref{eq:probucond} and
\eqref{eq:taucond}, respectively. Using these
conditional distributions, we develop the following full Gibbs
sampler for Bayesian probit mixed models.
\begin{algorithm}[H]
\caption{The $n$th iteration of the full Gibbs sampler}
\begin{algorithmic}[1]
  \label{alg:bprobifg}
  \STATE Draw $\lambda_{j}^{(n)} \overset{ind}\sim$ Gamma$(a_{j} + q_{j}/2, b_{j} + u_{j}^\top u_{j}/2), \,j = 1,\dots,r$ with $u = u^{(n-1)}$. \STATE Draw $v_i \overset{\text{ind}}\sim \text{TN}(\gamma_i,1,y_i), \, i=1,\dots, m,$ with $u = u^{(n-1)}$ and $\beta =\beta^{(n-1)}$.
\STATE Draw $u^{(n)} \sim \eqref{eq:probucond}$  with $\lambda = \lambda^{(n)}$ and $v = v^{(n)}$. 
\STATE Draw $\beta^{(n)} \sim \eqref{eq:probbetcond}$ with $v = v^{(n)}$ and $u = u^{(n)}$.
\end{algorithmic}
\end{algorithm}

It is known that blocking parameters can improve the performance of a
Gibbs sampler in terms of reducing its operator norm
\citep{liu:wong:kong:1994}. When one or more variables are correlated,
sampling them jointly can generally improve efficiency of the MCMC
algorithms. On the other hand, blocking may result in complex
conditional distributions that are not easy to sample from. For the
probit linear mixed models, \cite{wang:roy:2018} show that an
efficient two-block Gibbs sampler can be constructed by using the two
blocks, $\eta \equiv (\beta^\top,u^\top)^\top$ and
$(v,\lambda)$. Below we present \pcite{wang:roy:2018} block Gibbs
sampler.

Let $E = (X,Z)$ with the $i^{th}$ row being
$e_{i}^\top$ for $i=1,\dots,n$. Thus,
$\gamma_i = x_{i}^\top\beta + z_{i}^\top u = e_{i}^\top \eta$. From
\eqref{eq:bprobjt}, we have the conditional density of $\eta$ given $\lambda, v, y$ as
\begin{align}
  \label{eq:probetacond}
  f(\eta|\lambda, v, y) &\propto \left[ \prod_{i=1}^m  \exp\left[-\frac{1}{2}\left(v_i-x_i^{\top} \beta - z_i^{\top} u\right)^2\right]\right] \left[\prod_{j=1}^{r}\exp\left[-\frac{1}{2}\lambda_j u_j^{\top}u_j\right]\right] \nonumber\\ &\times \exp\left[-\frac{1}{2}\left(\beta-\mu_{0}\right)^{\top}Q\left(\beta-\mu_{0}\right)\right]\nonumber\\ & \propto  \exp\left[-\frac{1}{2}\left(v-E \eta\right)^{\top}\left(v-E\eta\right)\right] \cdot\exp\left[-\frac{1}{2} \eta^{\top} A(\lambda) \eta + \eta^T \theta \right],
\end{align}
where
\begin{equation}
  \label{eq:batau}
  \theta_{(p+q)\times 1} = \begin{pmatrix} Q\mu_{0} \\ 0_{q \times 1} \end{pmatrix}\; \mbox{and}\; A(\lambda)_{(p+q)(p+q)} = \begin{pmatrix}   
                        Q  & 0 \\
                         0  & D(\lambda)
                       \end{pmatrix}.
\end{equation}
Thus, from \eqref{eq:probetacond} we have
\begin{equation}
  \label{eq:probetacondis}
\eta|\lambda, v, y \sim N_{p+q} ((E^\top E + A(\lambda))^{-1} (E^\top v+\theta),(E^\top E + A(\lambda))^{-1}).  
\end{equation}
From \eqref{eq:bprobjt} note that conditional on $(\eta, y)$, $v$ and
$\lambda$ are independent. Thus, \eqref{eq:probetacondis} together with
\eqref{eq:problatcond} and \eqref{eq:taucond} result in the following
two-block Gibbs sampler for exploring the joint density \eqref{eq:bprobjt}.
\begin{algorithm}[H]
  \caption{The $n$th iteration for the two block Gibbs sampler}
  \label{alg:probbg}
  \begin{algorithmic}[1]
  \STATE  Draw $\lambda_{j}^{(n)} \overset{ind}\sim$ Gamma$(a_{j} + q_{j}/2, b_{j} + u_{j}^\top u_{j}/2), \,j = 1,\dots,r$ with $u = u^{(n-1)}$
    , and independently draw $v_i^{(n)}| \eta^{(n-1)}, y \overset{\text{ind}}\sim \text{TN}(e_i^{\top} \eta^{(n-1)},1,y_i)$ for  $i=1,\dots, m$.
%\STATE[1]
%\hspace{.18in}
\STATE Draw $\eta^{(n)}$ from \eqref{eq:probetacondis}, that is,
\[\eta^{(n)} \sim N\left(\left(E^{\top}E+ A(\lambda^{(n)})\right)^{-1}\left(E^{\top}v^{(n)}+\theta\right),\left(E^{\top}E+ A(\lambda^{(n)})\right)^{-1}\right).\]
\end{algorithmic}
\end{algorithm}

As in Section~\ref{sec:acda}, we now construct a Haar PX-DA algorithm
improving the block Gibbs sampler (Algorithm~\ref{alg:probbg}). In
order to construct the extra step in the Haar PX-DA, we derive the
marginal posterior density of $(\lambda, v)$ from the joint density
\eqref{eq:bprobjt} as
\begin{eqnarray}
\label{eq:margvtau}
  f\left(\lambda, v |y \right) & = & \int_{\mathbb{R}^{p+q}}f(\eta, \lambda, v | y) d\eta\\
                                           & \propto & \prod_{i=1}^{m}\left[1_{\text{(0,\ensuremath{\infty})}}\left(v_{i}\right)\right]^{y_{i}}\left[1_{\left(-\infty,0\right]}\left(v_{i}\right)\right]^{1-y_{i}} \prod_{j=1}^{r}\lambda_{j}^{\frac{q_{j}}{2}}\nonumber\\
                                           & \times & \exp\left\{ -\frac{1}{2}\left[v^{\top}E_1v-2v^{\top}E_2 - \theta^{\top}(E^\top E + A(\lambda))^{-1}\theta\right]\right\} \prod_{j=1}^r f(\lambda_j), \nonumber
\end{eqnarray}
where
\[
E_1 = \left[I_m-E\left(E^\top E + A(\lambda)\right)^{-1}E^{\top}\right]\;\;\mbox{and}\;\; E_2 = E\left(E^\top E + A(\lambda)\right)^{-1}\theta.
\]
For constructing an efficient sandwich step, \cite{wang:roy:2018} let the
group $\psi$ act on $\mathcal{V} \times \mathbb{R}_+^r$ through a
group action
$T^*(v, \lambda) = (hv,\lambda) = (hv_1, hv_2, \dots,
hv_m, \lambda)$. With the group action defined this way, it can be
shown that the Lebesgue measure on $\mathcal{V} \times \mathbb{R}_+^r$
is relatively left invariant with multiplier $\chi (h) = h^m$
\citep{roy:2014, hobe:marc:2008}. \cite{wang:roy:2018} then
consider a probability density function $\omega^*(h)$ on $\psi$ where
\begin{eqnarray}
\label{eq:g}
\omega^*\left(h\right) dh &\propto&  \nonumber f\left(\lambda, hv| y\right) \chi \left( h \right) \nu(dh)\\
& \propto & h^{m-1}\exp\left\{ -\frac{1}{2}\left[h^{2}v^{\top} E_1 v -  2hv^{\top} E_2 \right]\right\} dh.
\end{eqnarray}
Given $(\lambda, v)$, $\omega^*\left(h\right)$ is a valid density since
$E_1$ is a positive definite matrix. Below are the three steps involved in every
iterations of \pcite{wang:roy:2018} Haar PX-DA algorithm.

\begin{algorithm}[H]
\caption{The $n$th iteration for the Haar PX-DA algorithm}
\label{algorithm_g}
\begin{algorithmic}[1]
  \STATE  Draw $\lambda_{j}^{(n)} \overset{ind}\sim$ Gamma$(a_{j} + q_{j}/2, b_{j} + u_{j}^\top u_{j}/2), \,j = 1,\dots,r$ with $u = u^{(n-1)}$, and independently draw $v_i^{(n)}| \eta^{(n-1)}, y \overset{\text{ind}}\sim \text{TN}(e_i^{\top} \eta^{(n-1)},1,y_i)$ for  $i=1,\dots, m$.
%\STATE[1]
%\hspace{.18in}
  
\STATE Draw $h$ from \eqref{eq:g}.

\STATE Calculate
  $v_i^{\prime} = hv_i$ for $i=1,\dots,m$, and draw $\eta^{(n)}$ from \eqref{eq:probetacondis} conditional on
  $v^{\prime} = (v_1^{\prime},\dots, v_m^{\prime})^{\top}$, that is, draw
\[\eta^{(n)} \sim N\left(\left(E^{\top}E+ A(\lambda^{(n)})\right)^{-1}\left(E^{\top}v'+\theta\right),\left(E^{\top}E+ A(\lambda^{(n)})\right)^{-1}\right).\]
 \end{algorithmic}
\end{algorithm}

The adaptive rejection sampling algorithm \citep{gilk:wild:1992} can
be used to sample from $\omega^*(h)$ as it
log-concave. % The only difference between the Haar PX-DA
% algorithm~\ref{algorithm_g} and the block Gibbs sampler
% (Algorithm~\ref{alg:probbg}) is a single draw from the univariate
% density $q(g)$, which is easy to sample from. Thus, the computational
% burden, per iteration, for the Haar PX-DA algorithm is similar to that
% of the block Gibbs sampler. On the other hand, i
In general, the PX-DA algorithm is known to be theoretically more
efficient than the DA \citep{hobe:marc:2008, roy:2012a}. In the
context of some numerical examples of the probit GLMs,
\cite{roy:hobe:2007} showed that huge gains in efficiency are possible
by using the Haar PX-DA algorithm instead of the DA algorithm of
\cite{albe:chib:1993} \cite[see also][for comparisons of DA and PX-DA
algorithms for the robit GLM]{roy:2012b, roy:2014}.  A numerical
comparison of the three samplers presented in this section for the probit mixed models is given
in Section~\ref{sec:num}.
% A numerical comparison of the block Gibbs
% sampler (Algorithm~\ref{alg:probbg}) and the Haar PX-DA algorithm
% (Algorithm~\ref{algorithm_g}) for the probit mixed models can be found in
% an earlier version of \cite{wang:roy:2018} (available on the arXiv
% link of the paper).

\subsubsection{Data augmentation for Bayesian logistic mixed models}
\label{sec:bpgda}
Define the joint posterior density of $u,\beta,\lambda,w$ given $y$ as 
\begin{align}
\label{eq:blogijt}
f(u,\beta,\lambda,w \mid y) &\propto  \Big[\prod_{i=1}^{m}  \exp\{\kappa_{i} \gamma_i -w_{i}\gamma_i^{2}/2\}p(w_{i}) \Big] \phi_{q}(u;0,D(\lambda)^{-1}) f(\beta)f(\lambda)\nonumber\\
& = \Bigg[\prod_{i=1}^{m}  \exp\{k_{i}(x_{i}^\top \beta + z_{i}^\top u)-w_{i}(x_{i}^\top \beta + z_{i}^\top u)^{2}/2\} p(w_{i})\Bigg] \nonumber\\
& \times \phi_{q}(u;0,D(\lambda)^{-1}) \times  \prod_{j=1}^{r} \lambda_{j}^{a_{j}-1}\exp(-b_{j}\lambda_{j})\nonumber\\ & \times\exp\Big[-\frac{1}{2}(\beta-\mu_{0})^\top Q(\beta-\mu_{0})\Big],
\end{align}
where \eqref{eq:blogijt} follows from the priors on $\beta$ and
$\lambda$ given in \eqref{eq:betaprior} and \eqref{eq:tauprior},
respectively. From \eqref{eq:pglat} it follows that the
$(u, \beta,\lambda)-$ marginal of \eqref{eq:blogijt} is the target
density \eqref{eq:jointpd}.

Based on \eqref{eq:blogijt}, as in \cite{rao:roy:2021}, the conditional density of $\beta$ given $u, \lambda, w, y$ is 
\begin{align*}
f(\beta \mid  u, \lambda, w, y) &\propto \prod_{i=1}^{m} \exp\big[\kappa_{i} x_{i}^\top \beta-w_{i}(x_{i}^\top \beta)^{2}/2 - w_{i}(x_{i}^\top \beta) (z_{i}^\top u) \big]\\ & \times \exp\Big[-\frac{1}{2}(\beta-\mu_{0})^\top Q(\beta-\mu_{0})\Big] \\
&\propto \exp \Big[ -\frac{1}{2}\beta^\top (X^\top W X + Q)\beta+ \beta^\top (X^\top \kappa + Q \mu_{0} - X^\top W Z u) \Big].
\end{align*}
Thus, the conditional distribution of $\beta$ given $u, \lambda, w, y$ is 
\begin{align}
\label{eq:logibetcond}
\beta \mid u,\lambda, w, y \sim N((X^\top W X + Q)^{-1} (X^\top \kappa + Q \mu_{0} - X^\top W Z u),(X^\top W X+ Q)^{-1}).
\end{align}
The conditional densities of $w, u$ and $\lambda$ are given in
\eqref{eq:distributionomega}, \eqref{eq:logiucond} and
\eqref{eq:taucond}, respectively. \cite{rao:roy:2021} use these
conditional distributions to construct the following full Gibbs
sampler for Bayesian logistic mixed models.
\begin{algorithm}[H]
\caption{The $n$th iteration of the full Gibbs sampler}
\begin{algorithmic}[1]
  \label{alg:blogifg}
  \STATE Draw $\lambda_{j}^{(n)} \overset{ind}\sim$ Gamma$(a_{j} + q_{j}/2, b_{j} + u_{j}^\top u_{j}/2), \,j = 1,\dots,r$ with $u = u^{(n-1)}$. \STATE Draw $w_{i}^{(n)} \overset{ind}{\sim} PG(\ell_i, \gamma_i),\, i = 1,\dots,m$ with $u = u^{(n-1)}$ and $\beta =\beta^{(n-1)}$.
\STATE Draw $u^{(n)} \sim \eqref{eq:logiucond}$  with $\lambda = \lambda^{(n)}$, $\beta = \beta^{(n-1)}$ and $w = w^{(n)}$. 
\STATE Draw $\beta^{(n)} \sim \eqref{eq:logibetcond}$ with $w = w^{(n)}$ and $u = u^{(n)}$.
\end{algorithmic}
\end{algorithm}

\cite{rao:roy:2021} also construct an efficient two-block Gibbs sampler
with the two blocks being $\eta$ and
$(w,\lambda)$. From
\eqref{eq:blogijt}, 
the conditional density of $\eta$ given $\lambda,w,y$ is given by
\begin{align*}
f(\eta \mid \lambda,w,y) &\propto \prod_{i=1}^{n} \exp\big[\kappa_{i}e_{i}^\top \eta-w_{i}(e_{i}^\top \eta)^{2}/2\big]
\exp\big[-u^\top D(\lambda) u/2\big]\\ & \times \exp\big[-(\beta-\mu_{0})^\top Q( \beta-\mu_{0})/2\big].
\end{align*}
Thus, the conditional distribution of $\eta$ given
$\lambda,w,y$ is given by
\begin{align}
\label{eq:logietacond}
\eta \mid \lambda,w,y \sim N((E^\top W E + A(\lambda))^{-1} (E^\top\kappa+\theta),(E^\top W E + A(\lambda))^{-1}),
\end{align}
where $A(\lambda)$ and $\theta$ are defined in \eqref{eq:batau}.

      \eqref{eq:logietacond} together with \eqref{eq:distributionomega} and
\eqref{eq:taucond} result in the following two-block Gibbs sampler.                 
\begin{algorithm}[H]
\caption{The $n$th iteration of the block Gibbs sampler}
\begin{algorithmic}[1]
  \label{alg:blogibg}
  \STATE Draw $\lambda_{j}^{(n)} \overset{ind}\sim$ Gamma$(a_{j} + q_{j}/2, b_{j} + u_{j}^\top u_{j}/2), \,j = 1,\dots,r$ with $u = u^{(n-1)}$, and independently draw $w_{i}^{(n)} \overset{ind}{\sim} PG(\ell_i,e_{i}^\top\eta^{(n-1)}),\, i = 1,\dots,m$.
\STATE Draw $\eta^{(n)} \sim \eqref{eq:logietacond}$ with $\lambda = \lambda^{(n)}$ and $w = w^{(n)}$.
\end{algorithmic}
\end{algorithm}
A comparison of the full Gibbs sampler (Algorithm~\ref{alg:blogifg}) and
the block Gibbs sampler (Algorithm~\ref{alg:blogifg}) in the context
of some numerical examples as the dimensions of the design matrices
vary can be found in \cite{rao:roy:2021}.

% \section{Poisson GLMM}
% \label{sec:pois}

% %https://bbolker.github.io/mixedmodels-misc/ecostats_chap.html

% {\bf Comparison with MCMCglmm ?}

\section{Numerical example}
\label{sec:num}
In this section, we consider a publicly available simulated data set
named ``pbDat'' from the R package pbnm to compare the full Gibbs
sampler (Algorithm~\ref{alg:bprobifg}), the block Gibbs sampler
(Algorithm~\ref{alg:probbg}) and the Haar PX-DA algorithm
(Algorithm~\ref{algorithm_g}) for the probit mixed models.  This data
set has $m=100$ binary observations. There are $p=3$ covariates
including an intercept term. There is $r=1$ random effect with
$q_{1} = 12$ levels. We analyze the data set by fitting probit linear
mixed models with a normal prior \eqref{eq:betaprior} on $\beta$ with
$\mu_{0} = 0 $ and $Q =0.001 \I_{3} $ and a Gamma prior
\eqref{eq:tauprior} on $\lambda_{1}$ with $a_{1} =0.01$ and
$b_{1} = 0.01$. We ran the three samplers for $N=100,000$ iterations
starting at an initial value $(\beta^{(0)}, u^{(0)})$ with burn-in
$B=20,000$ iterations. Here $\beta^{(0)}$ is the estimate of $\beta$
obtained by fitting a probit linear model without any random
effect. The initial value $u^{(0)}$ is a sample drawn from
$N(0,(1/\lambda_{1}^{(0)})\I_{12})$ where $1/\lambda_{1}^{(0)}$ is the
estimate of random effect variance component obtained from the R
package lme4. We use the R package ars to make draws from the density \eqref{eq:g}.  

Next, we compare the performance of the full Gibbs (FG) sampler, the
block Gibbs (BG) sampler and the Haar PX-DA algorithm in the context
of this pbDat data.  The samplers are compared using lag $k$
autocorrelation function (ACF) values $k = 1,...,5$, effective sample
size (ESS) and multivariate ESS (mESS) (See \cite{roy:2020} for a
simple introduction to the different convergence diagnostic
measures.).  The ESS and mESS are calculated using the R package
mcmcse. We also compute the mean squared jumps (MSJ) (defined as
$\sum_{i=B+1}^{N} \norm{\beta^{(i+1)}-\beta^{(i)}}^2/(N-B)$ for the
$\beta$ variable, and similarly for the other variables. Here,
$\norm{\cdot}$ denotes the Euclidean norm. Lower ACF values and higher
ESS and MSJ numbers are preferred. Table~\ref{tab:simacf} provides
the values of ACF for the three samplers. Better performance of the
Haar PX-DA and the block Gibbs samplers compared to the full Gibbs
sampler is observed from their smaller ACF
values. Table~\ref{tab:simmess} provides the ESS values of the
intercept parameter, first two regression coefficients and
$\lambda_1$. It also gives the mESS values for $u$ and
$(\beta, \lambda_1)$. Again, better efficiency of the Haar PX-DA and
the block Gibbs samplers compared to the full Gibbs sampler is
demonstrated from their larger ESS and mESS values.  From
Table~\ref{tab:simmsj}, it can be seen that the Haar PX-DA sampler
leads to higher MSJ values than the full Gibbs sampler. Also, the
block Gibbs sampler results in higher MSJ values than the full Gibbs
sampler with the exception of $\lambda_{1}$. Thus,
Table~\ref{tab:simmsj} also corroborates better mixing of the Haar
PX-DA and the block Gibbs samplers than the full Gibbs sampler.
\begin{center}
\begin{table*}[h]
  \caption{ACF for different samplers for the pbDat data}
  \centering
\begin{tabular}{ccccccc}
\hline\hline Parameter&  Sampler  & lag 1& lag 2& lag 3& lag 4& lag 5\\
 \hline
  $\beta_{0}$&FG& 0.957 & 0.915 & 0.876& 0.838&0.801\\
                      &BG& 0.085& 0.062&0.042 & 0.023& 0.026\\
                      &Haar&0.089& 0.061& 0.038& 0.030 & 0.020\\
\hline         
$\beta_{1}$&FG&  0.724 & 0.539& 0.410 & 0.317& 0.249\\
                      &BG& 0.701& 0.510& 0.386& 0.297& 0.234\\
                      &Haar& 0.680& 0.472& 0.334& 0.243& 0.177\\
           \hline
  $\beta_{2}$&FG& 0.836& 0.715& 0.622& 0.547& 0.488\\
                      &BG& 0.812& 0.681& 0.584& 0.509& 0.449\\
                      &Haar& 0.720& 0.528& 0.395& 0.297& 0.223\\
           \hline
  $\lambda_{1}$&FG& 0.682& 0.597& 0.549& 0.486&0.424\\
                      &BG&0.635& 0.521& 0.442& 0.375& 0.320\\
                      &Haar& 0.636& 0.442& 0.344& 0.266& 0.207\\
           \hline
\end{tabular}
\label{tab:simacf}
\end{table*}
\end{center}

\begin{center}
\begin{table*}[h]
  \caption{Multivariate ESS and ESS for different samplers for the pbDat data}
  \centering
\begin{tabular}{cccccccc}
\hline\hline  Sampler  & mESS ($\beta\; \lambda$)  &mESS ($u$)& ESS ($\beta_{0}$)&ESS ($\beta_{1}$)&ESS ($\beta_{2}$)&ESS ($\lambda_{1}$)\\
 \hline
FG&4915&14036&1565&8362&3481&4558\\
  BG&13142&19156&47437&8867&4449&7093\\
  Haar&18865&21940&47964&12891&10921&11177\\
\hline
\end{tabular}
\label{tab:simmess}
\end{table*}
\end{center}

\begin{center}
\begin{table*}[h]
  \caption{Mean squared jumps for different samplers for the pbDat data}
  \centering
\begin{tabular}{ccc|ccc|ccc}
\hline\hline  \multicolumn{3}{c}{FG}& \multicolumn{3}{c}{BG}& \multicolumn{3}{c}{Haar}\\
\hline
 $\beta$&$u$&$\lambda$& $\beta$&$u$&$\lambda$& $\beta$&$u$&$\lambda$\\
 \hline
\phantom{} \phantom{} 0.018&0.224&0.232&\phantom{} 0.154&0.613&0.198&\phantom{} 0.155&0.622&0.235\\
 \hline
\end{tabular}
\label{tab:simmsj}
\end{table*}
\end{center}
% \section{Binomial GLMMs on pbDat data}
% \label{sec:bin}

\section{Discussion}
\label{sec:disc}
In this article, we have presented several MCMC algorithms for both
frequentist as well as Bayesian GLMMs. While some of these algorithms
discussed here are available in the literature, others are developed
here. Since these algorithms result in Harris ergodic Markov chains,
the (Monte Carlo) sample averages are consistent estimators of the
means with respect to the corresponding target densities. On the other
hand, in practice, it is important to ascertain the errors associated
with these Monte Carlo estimates. An advantage of being able to
calculate a valid standard error is that it can be used to decide
`when to stop' running the MCMC chain \citep{roy:2020}. A valid
standard error for the Monte Carlo estimates can be formed if a
central limit theorem is available for the time average estimator.
Establishing geometric ergodicity (GE) of the underlying Markov chains
is the most standard method for guaranteeing a central limit theorem
for MCMC based estimators \citep{meyn:twee:1993}. GE is also used for
consistently estimating the asymptotic variance in the central limit
theorem \citep{vats:fleg:jone:2019}.

For several of the MCMC algorithms presented here, GE has been
established in the literature. For example, \cite{roy:zhan:2021}
demonstrates GE of Markov chains underlying different MALA for GLMMs. The GE of
the P\'{o}lya Gamma block Gibbs sampler for Bayesian logistic mixed models
under proper and improper priors have been established in
\cite{wang:roy:2018a} and \cite{rao:roy:2021}, respectively.
\cite{wang:roy:2018} derive conditions under which the block Gibbs
sampler and the Haar PX-DA algorithm for the probit mixed models are
geometrically ergodic when improper priors are assumed on the
regression coefficients and the variance components. 

It would be interesting to construct and study efficient DA samplers
for other GLMMs, for example the GLMMs with the robit link
\citep{roy:2012b}.  A potential future study can be to extend
\pcite{wang:roy:2018} GE results to the probit mixed models with
proper priors. Another potential project is to study convergence
properties of the HMC chains in the context of GLMMs.

%   Reference for

% Convergence properties

% Numerical examples
%future: DA for Poisson GLMs and GLMMs

\bibliographystyle{ims}
% \bibliography{C:/Users/vroy/Box/Misclen/LocalTexFiles/bibtex/bib/misc/refs}

\end{document}